\documentclass{caosp305}

\usepackage{graphicx}

\usepackage{natbib}
\articleNo{0}
\pubyear{2017}
\volume{0}
\volnumber{0}
\firstpage{0}
\received{Month 32, 2017}
\accepted{Month 33, 2017}

\def\BibTeX{{\rm B\kern-.05em{\sc i\kern-.025em b}\kern-.08em
             T\kern-.1667em\lower.7ex\hbox{E}\kern-.125emX}}

\begin{document}

%
\hauthor{M.\,Yu}

\title{Have we seen all glitches?}


%
\author{
        M.\,Yu \inst{1}
       }

%
\institute{ National Astronomical Observatories of China, Beijing
  100101 \newline \email{vela.yumeng@gmail.com} }


\maketitle

\begin{abstract}
Neutron star glitches are observed via artificially scheduled pulsar
pulse arrival-time observations. Detection probability density of
glitch events for a given data set is essentially required knowledge
for realizing glitch detectability with specified observing system
and schedule. In \citeauthor{yl17}, the detection probability density
was derived for the \citeauthor{ymh+13} data set. In this proceeding,
further discussions are presented.
\end{abstract}

Hundreds of pulsars have been being observed at the Parkes Observatory
over decades via a number of timing programmes. Search for pulse
frequency glitches for a data set that contains an entire 1911\,yr
observations over 165 pulsars was done by \citet{ymh+13}. The result
is 107 glitches were identified in 36 pulsars. For the \citet{ymh+13}
data set, glitch identification depends on the observing cadences and
the observed timing noise. \citet{yl17} thus defined the complete
probability formula of identifying glitch (see Eq. 1 therein). They
found the derived detection probability densities for both the group
and individual cases are not uniform; as glitch becomes larger the
density increases (see Figs. 5, 7 and 8 therein). The high-cadence
observations for the Crab pulsar showed the $\Delta\nu$ values of
glitches are significantly larger than those of timing noise
\citep{eas+14}. These imply the cadences with which \citet{ymh+13}
pulsars were observed were not adequate to observe all glitches
occurred.

\begin{figure}
\begin{center}
\includegraphics[width=11.5cm,angle=-90]{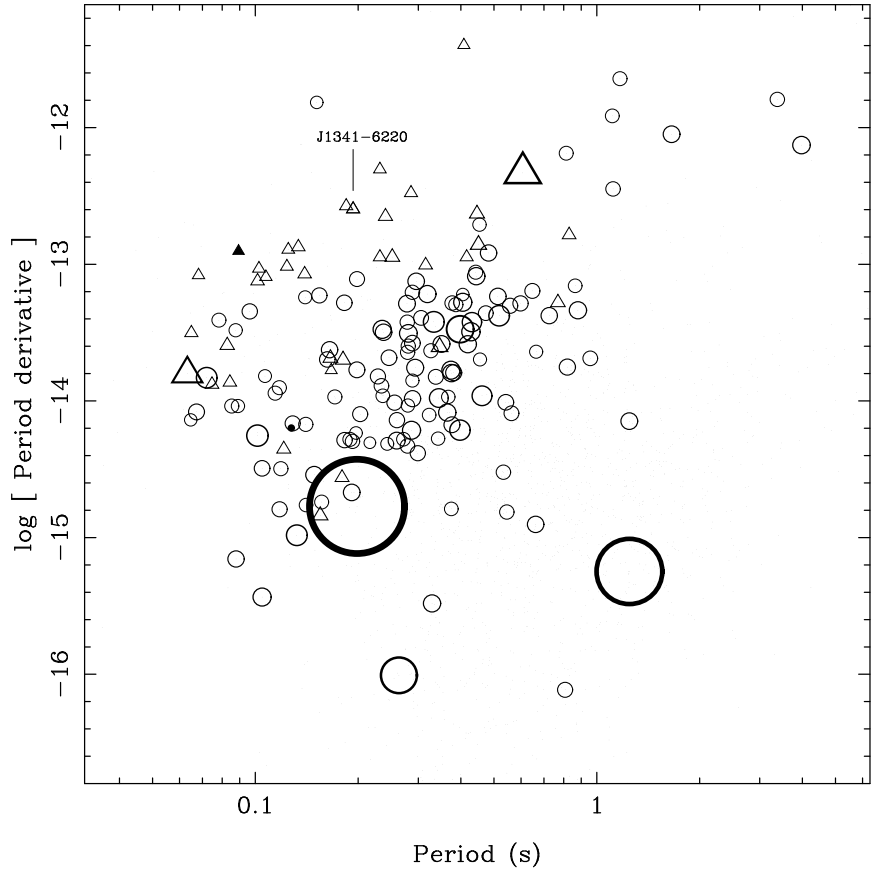}
\end{center}
\caption{Average ToA intervals on the $P - \dot{P}$ diagram. Circles
  indicate the non-glitching pulsars, triangles indicate the glitching
  pulsars. Minima are shown by solid symbols.}\label{fig:avg}
\end{figure}

\begin{figure}
\begin{center}
\includegraphics[width=11.5cm,angle=-90]{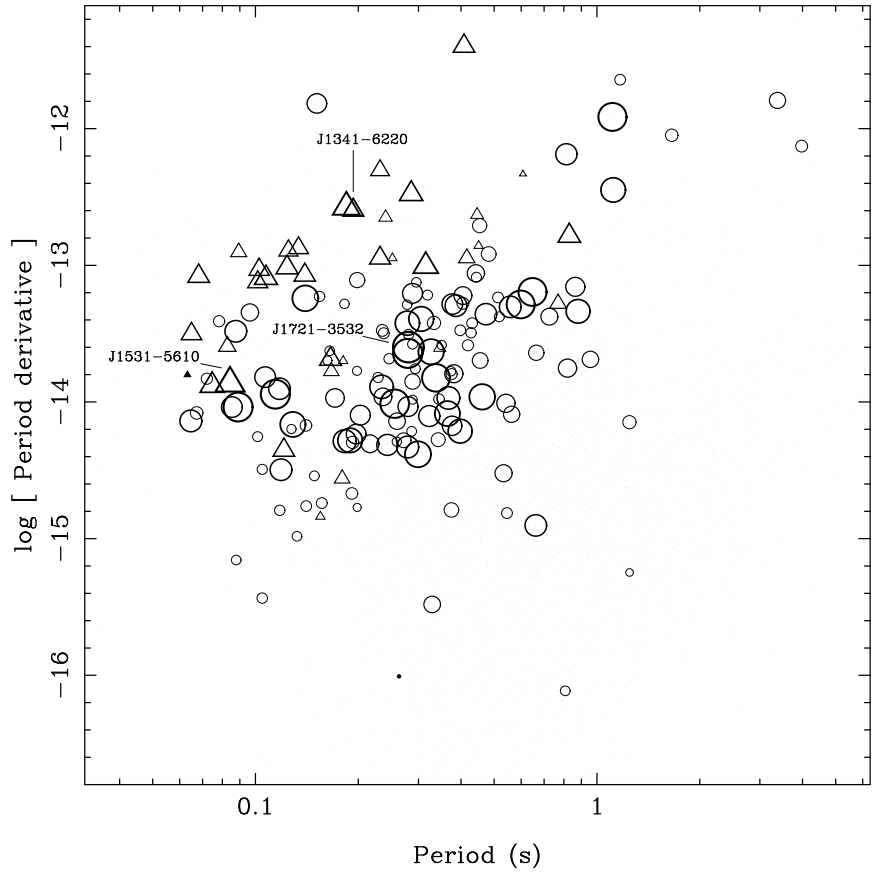}
\end{center}
\caption{ToA interval modulation index on the $P - \dot{P}$
  diagram. Circles indicate the non-glitching pulsars, triangles
  indicate the glitching pulsars. Minima are shown by solid
  symbols.}\label{fig:mod}
\end{figure}

\begin{figure}
\begin{center}
\includegraphics[width=11.5cm,angle=-90]{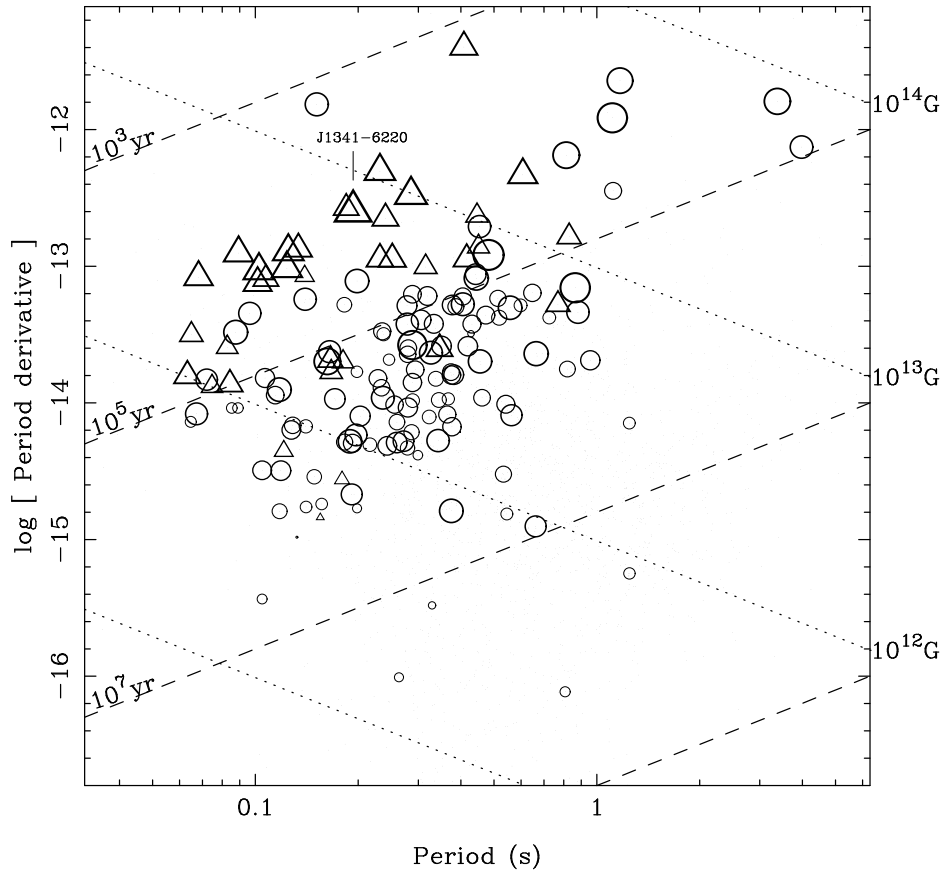}
\end{center}
\caption{Amplitude of power spectral density of the observed timing
  noise on the $P - \dot{P}$ diagram. Circles indicate the
  non-glitching pulsars, triangles indicate the glitching
  pulsars.}\label{fig:amp}
\end{figure}

In Fig. \ref{fig:avg}, average pulse time-of-arrival (ToA) interval
for each of the 165 pulsars is plotted on the $P - \dot{P}$
diagram. Symbol size is a linear function of the value with positive
slope. For the non-glitching pulsars, the averages range between
10.4\,d for PSR~J1359$-$6038 and 798.7\,d for PSR~J1047$-$6709; most
averages are a few tens of days. For the glitching pulsars, the
averages range between 8.3\,d for the Vela pulsar and 240.6\,d for
PSR~J1740$-$3015. For PSR~J1740$-$3015, the low observing cadences result in the
unidentification of thirteen glitches in the \citet{ymh+13}
data. PSR~J1341$-$6220 which shows seventeen glitches was
on average observed every 23.5\,d. In Fig. \ref{fig:mod}, ToA interval
modulation index or the average over the intervals' standard deviation
for each pulsar is shown. (Symbol size is a linear function of the
value with positive slope.) For the non-glitching pulsars, the indices
range between 0.08 for PSR~J1456$-$6843 and 1.20 for
PSR~J1721$-$3532. For the glitching pulsars, the indices range between
0.06 for PSR~J1105$-$6107 and 1.12 for
PSR~J1531$-$5610. PSR~J1341$-$6220 has 0.73. In Fig. \ref{fig:amp},
amplitude of power spectral density of observed timing noise is
shown. (Symbol size is a linear function of the value with positive
slope.) PSR~J1341$-$6220 has the maximum amplitude in the sample. In
general, young pulsars show large timing noise while timing noise does
not seem to correlate with magnetic field.

\newpage

\newpage

\acknowledgements
The work \citet{yl17} is supported by the National Natural Science
Foundation of China (No. 11403060), the Joint Research Fund in
Astronomy (U1531246) under cooperative agreement between the National
Natural Science Foundation and Chinese Academy of Sciences, the
Strategic Priority Research Program `The Emergence of Cosmological
Structures' of the Chinese Academy of Sciences (No. XDB09000000), the
International Partnership Program of the Chinese Academy of Sciences
(No. 114A11KYSB20160008) and the Strategic Priority Research Program
of the Chinese Academy of Sciences (No. XDB23000000).

\bibliography{myu}
\end{document}